\documentclass{article}

\usepackage{arxiv}

\usepackage[utf8]{inputenc} % allow utf-8 input
\usepackage[T1]{fontenc}    % use 8-bit T1 fonts
\usepackage{hyperref}       % hyperlinks
\usepackage{url}            % simple URL typesetting
\usepackage{booktabs}       % professional-quality tables
\usepackage{amsfonts}       % blackboard math symbols
\usepackage{nicefrac}       % compact symbols for 1/2, etc.
\usepackage{microtype}      % microtypography
\usepackage{lipsum}
\usepackage{graphicx}

\usepackage{authblk}
\usepackage[numbers]{natbib}
\usepackage{makecell}
\usepackage{soul}
\usepackage[font=bf]{caption}
\usepackage{enumitem}
\newcommand{\qt}[2]{
\begin{quote}
\textbf{#1}: \textit{#2}
\end{quote}
}

\newcommand{\qti}[2]{
\textit{``#2''} ({#1})
}
\graphicspath{ {./images/} }
\AtBeginDocument{%
  }

\title{Exploring the Requirements of Clinicians for Explainable AI Decision Support Systems in Intensive Care}

\author[1]{Jeffrey N. Clark}
\author[1]{Matthew Wragg}
\author[1]{Emily Nielsen}
\author[1]{Miquel Perello-Nieto}

\author[1]{Nawid Keshtmand}
\author[1, 2]{Michael Ambler}
\author[2]{Shiv Sharma}
\author[2]{Christopher P. Bourdeaux}
\author[1]{Amberly Brigden}
\author[1]{Raul Santos-Rodriguez}

\affil[1]{University of Bristol, UK}
\affil[2]{University Hospitals Bristol and Weston NHS Foundation Trust, UK}

\begin{document}
\maketitle
\begin{abstract}
There is a growing need to understand how digital systems can support clinical decision-making, particularly as artificial intelligence (AI) models become increasingly complex and less human-interpretable. This complexity raises concerns about trustworthiness, impacting safe and effective adoption of such technologies. Improved understanding of decision-making processes and requirements for explanations coming from decision support tools is a vital component in providing effective explainable solutions. This is particularly relevant in the data-intensive, fast-paced environments of intensive care units (ICUs). To explore these issues, group interviews were conducted with seven ICU clinicians, representing various roles and experience levels. Thematic analysis revealed three core themes: (T1) ICU decision-making relies on a wide range of factors, (T2) the complexity of patient state is challenging for shared decision-making, and (T3) requirements and capabilities of AI decision support systems. We include design recommendations from clinical input, providing insights to inform future AI systems for intensive care.
\end{abstract}

% keywords can be removed

\section{Introduction}
Intensive care units (ICUs) support patients with the highest medical care requirements in hospitals, and out of necessity, generate huge amounts of rich healthcare data \cite{celi2013big}.
AI systems are increasingly being developed to assimilate existing data sources and create models to support decisions, aiming to aid ICU clinicians in their work \cite{shillan2019use, mcwilliams2019towards}. Many of these systems are developed for safety-critical decisions, yet are not human interpretable. Hence, their deployment is met with justifiable scepticism amongst clinical teams \cite{Galsgaard2022Aiam}, and the implementation and adoption of such systems in clinical practice often fail \cite{Yildirim2024Saic}.

Explainable AI (xAI) aims to bridge the gap, enabling human users to understand decisions from these systems. A wide range of xAI tools have been developed \cite{saeed2023explainable}, largely tailored towards AI experts and practitioners. There is a lack of understanding of clinicians' requirements for such tools and it is unclear how these tools should be integrated into intensive care workflows \cite{wen2022application}. Additionally, legislation such as the recent EU's AI Act \cite{EUAIact} is paving the way for individuals to access explanations of the decisions generated by these automated systems \cite{Pavlidis2024Utbb}. Further work is required to ensure that xAI tools fit the needs of healthcare professionals within intensive care\cite{wen2022application}, both to support their acceptance and to maximise their effectiveness in explaining the outputs of AI decision support systems. Our work seeks to understand the requirements of ICU healthcare professionals in the design and implementation of xAI tools within AI decision support systems. We are guided by two research questions: 

\begin{enumerate}[label={\textbf{RQ\arabic*:}}]
    \item How do clinicians currently use data and decision support systems to formulate and explain decisions in intensive care?
    \item What requirements do clinicians have for the design and implementation of explainable AI decision support systems in intensive care?
\end{enumerate}

To address our research questions, we ran two 45-minute workshops with seven intensive care clinicians of various roles and levels of seniority. Thematic analysis of the transcripts revealed three themes spanning the factors which affect how clinicians make decisions in ICU, the communication of such decisions, and the requirements of AI decision support systems -- including xAI. In summary, our contributions are as follows:

\begin{enumerate}
    \item Insights into current clinical decision-making considerations and practices in intensive care.
    \item Opportunities and challenges surrounding the use of AI decision support systems in intensive care.
    \item Design considerations for xAI systems that reflect the highly contextual nature of intensive care decisions.
\end{enumerate}

This study lays the groundwork for AI systems which better serve the needs of users within ICU, with potential applications more broadly within healthcare. In doing so, systems with those considerations would improve a clinician's ability to interact with clinical decision support systems, influencing their adoption and enabling improved patient care.

%%%%%%%%%%%%%%%%%%%%%%%%%%%%
%%                        %%
%%      Related Work      %%
%%                        %%
%%%%%%%%%%%%%%%%%%%%%%%%%%%%
\section{Related Work}

% importance of AI & then xAI within the ICU
The ICU is a hectic environment, requiring optimal communication and constant recording of bio-signals and medical events. High quantities of data are concurrently generated and stored in computer systems, which are later analysed by clinicians to make -- often quick -- decisions. Multiple research studies have been proposed to facilitate the difficulties that arise in the ICU with data-driven solutions \cite{moazemi2023artificial, wen2022application, alowais2023revolutionizing}.
%In order to facilitate the evaluation of such systems, various institutions have made ICU data publicly available for research \cite{sauer2022systematic} (e.g. MIMIC-IV \cite{mimiciv_v3}, eICU Collaborative REsearch Database (eICU-CRD) \cite{pollard2018eicu}, Amsterdam University Mecical Center data base \cite{thoral2021sharing}, and High time-resolution intensive care unit dataset \cite{hyland2020early}).
Several works have explored the use of AI within the ICU \cite{saqib2023artificial, wen2022application}, which highlights that AI can ``enhance the accuracy, efficiency and cost-effectiveness of medical diagnosis'' \cite{alowais2023revolutionizing}. However, the explainability aspects of AI recommendations in the clinical setting are still in the early stages. Explainability has been recognised as a critical component in integrating AI with clinical decision-making \cite{Lee2023Understanding}, and is particularly important for the safety-critical nature of intensive care decisions. 
% xAI in ICU does not consider HCI component
Initial studies on the explainability of AI systems within the ICU \cite{moazemi2023artificial, wen2022application} focus on a range of different prediction tasks, predominantly readmission \cite{hu2024explainable, gonzalez2023improving, hu2022explainable} and mortality prediction \cite{gonzalez2021using, barnes2024towards} with promising results in terms of predictive performance and feature importance explanation. However, they usually rely on off-the-shelf and domain-agnostic xAI toolboxes \cite{JMLR:v21:19-1035,sokol2022fat-forensics}, using techniques such as Shapley Additive Explanations \cite{lundberg2017unified} and Local Interpretable Model-agnostic Explanations \cite{tulio2016why}, but they overlook the human interaction element. \citet{panigutti2023co} argues that participatory approaches are a vital component in developing xAI, as effective AI decision support requires explanations based on the end-users' needs.

Drawing from participatory design research for interventions in the ICU, several works provide insights on contextual challenges within the ICU, highlighting the cognitively demanding nature of the ICU \cite{Faiola2015, EiniPorat2022} and co-designing low-risk AI concepts to promote easier adoption \cite{Yildirim2024Saic}. However, these works focus on general system requirements, rather than information requirements for effective use of AI suggestions. \citet{Orth2024} consider general information requirements for intensive care decision-making, such as the need for patient history and comprehensible information, and explores the implications on AI. Our paper builds on this work but focuses on system requirements for AI explanations by exploring the information requirements of ICU clinicians.

%%%%%%%%%%%%%%%%%%%%%%%%%%%%%%%%%%%%%%%
%%                                   %%
%%      Expert Interview Study       %%
%%                                   %%
%%%%%%%%%%%%%%%%%%%%%%%%%%%%%%%%%%%%%%%
\section{Expert Interview Study}

To better understand decision-based workflows in intensive care, and to gather clinicians' perspectives on both current decision-making processes and requirements of xAI decision support tools, we conducted semi-structured interviews with clinicians working in the ICU at a large teaching hospital in Bristol, UK. We invited seven clinicians working in several different roles within the ICU to participate in one of two semi-structured group interview sessions. Both interview sessions were conducted virtually using Microsoft Teams. Participants were recruited through word-of-mouth using expert sampling, to cover as wide a range of clinical roles as possible within the ICU. These roles included an anaesthetist trainee, intensive therapy nurse, and outreach nurse. No consultants were recruited due to staffing issues, as acknowledged in the limitations section. To preserve their anonymity, we refer to them by their general roles: Resident Doctors (\textbf{D}) and Nurses (\textbf{N}). Each participant provided their informed consent before the start of the study\footnote{Ethical approval for this project was granted by a Research Ethics Committee (REC) within the UK National Health Service (NHS) (IRAS 297886).}.

\subsection{Structure of the Interview Sessions}

At the beginning of the session, all seven participants introduced themselves, before being introduced to the four study facilitators, and were briefed on how the interview sessions would be conducted. Following the briefing, participants were split into two groups for the interview sessions, which ran in parallel using separate breakout rooms on Microsoft Teams. Participants \textbf{N1, N2, D1} and \textbf{D2} were assigned to \textbf{Room 1}, and \textbf{N3, N4} and \textbf{D3} to \textbf{Room 2}. Before the interviews, participants were presented with two example patient personas (\autoref{app:personas}), representing patients with differing symptoms and care requirements. These personas were created to provide participants with concrete examples they could use to illustrate or contextualise any points they wished to make. Participants were also provided with a shared digital whiteboard for each session to support discussions. Each session was guided by two facilitators and lasted approximately 35 minutes. Audio recordings of both interview sessions were taken, transcribed verbatim and fully anonymised by the authors present in each respective session. The sessions were semi-structured around two sets of questions, allowing participants to freely discuss their thoughts and experiences whilst retaining a focus on our research questions. These sets of questions are described below, with a full list of questions included in \autoref{app:questions}:

\begin{enumerate}[label=(\alph*)]
    \item \textit{Background Knowledge Gathering} (15 mins): The first set of interview questions focused on understanding clinicians' experience with making decisions in the ICU. This included information they find useful in different contexts, and how they communicate with colleagues around decisions.
    \smallskip
    \item \textit{Requirements of xAI Systems} (20 mins): The second set of interview questions focused on understanding clinicians' requirements of explainable AI (xAI) systems in the ICU, including what information they would like to see included in explanations, and how they would like these explanations to be presented to them.
\end{enumerate}

\subsection{Analysis}

Analysis of the interview transcripts was performed by the four facilitators using reflexive thematic analysis \citep{BraunClarke2018ThematicAnalysis}. A primarily inductive approach was taken, with codes generated through an open-coding process for text related to our research questions. For each group interview session, two authors independently coded the transcripts, before iteratively developing and agreeing on a shared code book. Once code books had been created for each session, the four facilitators collectively analysed and refined all codes from both sessions into a single combined code book. Following the synthesis of the combined code book, the four facilitators and an additional AI researcher collaboratively and iteratively generated themes from the codes and quotes. The resulting themes were reviewed to ensure they provided an accurate and complete representation of the data, and that they aligned with our research questions.

%%%%%%%%%%%%%%%%%%%%%%%%
%%                    %%
%%      Results       %%
%%                    %%
%%%%%%%%%%%%%%%%%%%%%%%%
\section{Results}

Through thematic analysis, we identified three themes pertinent to our research questions (\autoref{tab:themes_expanded}). In this section, we present details of each theme, substantiated by quotes taken from the interview transcripts. Some of the quotes we present have been edited to increase readability. This includes removing unintentionally repeated words and utterances such as ``umm'', and adding context to sentences where necessary using \textit{[square brackets]}. While these quotes differ slightly from the raw transcript, the context and meaning of each quote has been preserved.

\renewcommand{\arraystretch}{1.1}
\begin{table}[tb!]
    \small
    \centering
    \setlength{\belowcaptionskip}{5pt}
    \begin{tabular}{ l }
    \hline
    \textbf{T1. ICU Decision-Making Relies on a Wide Range of Factors}\\
    \quad \textbf{T1.1 Patient Journey:} Care planning, patient trajectory, history, and recent data\\
    \quad \textbf{T1.2 Patient Physiological Data:} Organ systems, diagnosis-specific data, variability in baseline, and trends \\
    \quad \textbf{T1.3 Importance of ``End-of-Bed'' Information:} Forming a complete picture of patients (e.g., behavioural \& visual factors)\\
    \quad \textbf{T1.4 Operational Factors:}  Bed and staff availability, hospital flow and resource constraints \\
    \hline

    \textbf{T2. Complexity of Patient State is Challenging for Shared Decision-Making} \\
    \quad \textbf{T2.1 Collaborative Decision-Making:} Checking decisions with colleagues, prompting investigation, \& family involvement \\
    \quad \textbf{T2.2 Patient Handover Between Clinicians:} Efficiency of handovers, and frameworks for information transfer \\
    \hline
    
    \textbf{T3. Requirements and Capabilities of AI Decision Support Systems}  \\
    \quad \textbf{T3.1 Framing and Presentation of Decision Recommendations:} Importance of evidencing recommendations, \\
    \qquad and framing for acceptance and trust \\

    \quad \textbf{T3.2 Complexity and Detail of Explanations:} Layered system, minimising cognitive load and number of systems \\
    \quad \textbf{T3.3 Impact of xAI Recommendations on Decision-Making:} Bias/obscure judgement, prompt new thought,\\ 
    \qquad and highlight missed information \\
    \hline
    \end{tabular}
    \normalsize
    \caption{Themes, sub-themes and key talking points identified from the interviews}
    \label{tab:themes_expanded}
\end{table}

%%%%%%%%%%%%%%%%%%%%%%%%%%%%%%%%%%%%%%%%%%%%%%%%%%%%
%% Theme 1: Factors Affecting ICU Decision-Making %%
%%%%%%%%%%%%%%%%%%%%%%%%%%%%%%%%%%%%%%%%%%%%%%%%%%%%
\subsection{ICU Decision-Making Relies on a Wide Range of Factors}
Intensive care decision-making is highly complex, with several important factors to consider. Beyond the expected physiological data, it was important to consider the holistic journey of the patient, as well as contextual factors from observing the patient's visual and behavioural state. In addition, it was important to consider wider constraints and operational factors, including the availability of beds and staffing.

\subsubsection{Patient Journey.} 

A patient's past, present, and trajectory towards a likely future state all form part of the decision-making process. Five of the participants raised that care planning and understanding a patient's journey are crucial in guiding treatment decisions, especially when considering escalation or comfort care. \textit{``Escalation planning''} was brought up by N1, N2, and N3, with N1 emphasising the need for it in cases of deterioration:  \textit{``would they need more ventilation\ldots etcetera''}. N4 stressed the importance of establishing \textit{``a direction of travel''}, whether it be full treatment or a palliative approach because \textit{``they’re two very different platters of options''}. Patient history also plays a vital role in decision-making, as N3 noted, where past experiences influence discharge timing: \textit{``a patient who's been to intensive care three or four times previous to this and has keep coming back from the ward back to intensive care\ldots we're a bit more reluctant to send them as soon as those vital numbers are correct, as we might do with another patient who's maybe postoperative for the first time''} (N3).

The significance of recent data and the patient’s current trajectory were raised. N1 highlighted the importance of flagging potential issues, for example \textit{``a recent blood result\ldots [such as] a really raised potassium''}, to reassess decision-making, and that \textit{``the trajectory of [a patient's] long-term illness''} will factor into clinical decisions. D3 pointed out that \textit{``it really depends on the context and causes of the deterioration''}, which informs whether further intervention is appropriate, adding that \textit{``everyone wants to make sure it's the right choice having seen that the patient has remained stable for at least 24 hours''}.

\subsubsection{Patient Physiological Data.}
Monitoring patient physiological data plays a crucial role in clinical decisions such as readiness for discharge, with participants highlighting the need for patients to meet specific physiological requirements. N3 explained that \textit{``the numbers being right are kind of stage 1 because otherwise they'll leave the unit and [immediately get flagged for readmission]''}. Which measurements are most important can be specific to the diagnosis, as D1 noted: \textit{`` every patient comes with a different pathology and therefore we're going to be looking at slightly different parameters for each one''}. However, even for the same pathology, what is considered a normal value varies between patients: \textit{``So, you say, okay, a normal blood pressure is 120 / 80, except that for certain groups of patients, that's absolutely abnormal.''} (N4). 

Categorising data by organ systems was discussed, with D2 suggesting that this could help track complex cases, as \textit{``pretty much every problem you can categorise as one of those [organ] systems''}. D2 went on to say: \textit{``I often find myself going through the charts kind of totting up organ systems and totting up problems to kind of explain to myself what's happening with a patient.''} (D2). Participants brought up the importance of looking at data trends rather than isolated data points. As noted by D3: \textit{``it's all about the trend\ldots has it improved''}. \textit{``There's obviously a lot of data that we collect and sometimes it's easier to see that as a trend''} D1 stated, and noted specifically, the importance of trends \textit{``getting or approaching ready for discharge''} (D1). This approach helps clinicians make informed decisions regarding patient progression.

\subsubsection{Importance of End-of-Bed Information.}
While physiological data is critical, participants emphasised that it is often insufficient to form a complete picture of a patient's condition. D2 highlighted that \textit{“you need to stand at the end of the bed, [and] look at the patient”} to assess whether the patient can be safely managed outside of intensive care. Behavioural factors were deemed to be difficult to capture in data, as D2 noted using the example of patients experiencing delirium, \textit{``they're needing constant \ldots physical and verbal intervention to manage their behaviour and keep them safe''}. Visual assessments, such as inviting senior staff to \textit{``eyeball''} the patient (N2) and assess \textit{``how the patient looks clinically''} (N3), were seen as invaluable for decision planning. This was echoed by D3, who explained that seeing the patient over time can provide reassurance by providing information \textit{``you can get from end of bed by just looking at the patient and assessing them that you can't necessarily from the computer''}.

\subsubsection{Operational Factors.}

As part of their decision-making process, clinicians must consider the wider hospital system:

\qt{N2}{``[Hospital] flow is generally a real big problem for us \ldots so finding that appropriate ward is difficult. \ldots [We must] make sure that they're going to get that appropriate step down of care when they leave us, which does often result in a delay in discharge to our wards.''}

Resource constraints and bed availability outside of the ICU are important factors when deciding upon discharge. Specifically, whether \textit{``it's an appropriate bed for this patient''} (D2) \textit{``on an appropriately staffed ward''} (D2). The consequences of this is clear: \textit{``we see daily that patients stay with us a bit longer because there aren't beds in the appropriate facility for them''} (N2). Considerations around \textit{``what time of the week it is''} and resulting staff availability were raised, concerning \textit{``what kind of medical and nursing cover they'll get in that in that discharge destination''} (D2). In particular: \textit{``If they're requiring specialist support from teams, are those teams available out of hours or on the weekends?''} (N1).

%%%%%%%%%%%%%%%%%%%%%%%%%%%%%%%%%%%%%%%%%%%%%%%%%%%%%%%%%%%%%%%%%
%% Theme 2: Communication of Decisions and Patient Information %%
%%%%%%%%%%%%%%%%%%%%%%%%%%%%%%%%%%%%%%%%%%%%%%%%%%%%%%%%%%%%%%%%%
\subsection{Complexity of Patient State is Challenging for Shared Decision-Making}
The complexity and numerous factors which inform decisions, as highlighted in the previous theme, presents a significant challenge with shared decision-making. However, collaborations were seen as important to check and question decisions, in addition, the decision-maker changes frequently due to shifts and working patterns. As such, communicating complex information effectively is highly important to facilitate familiarisation and decision support.
%clear overview of patient state important for communication.... difficulty of sharing information... multiple stakeholders... complexity of data and sharing info
\subsubsection{Collaborative Decision-Making.}
Participants valued collaborating with colleagues to \textit{``double check''} (D3) their decisions, especially where they could have missed patient information:

\qt{D2}{``we all miss things in the data and our patients’ pathways. There are times when you think you've made a decision and then someone points out a drug that was given or a blood test that's abnormal.''}

This may be due to the vast amounts of data, and the need for context-specific approaches to interpret them, which may make collaborative decision-making more challenging: \textit{``we always have lots of data and we deal with lots of abnormal data \ldots we're very used to accepting certain things in the clinical context.''} (D2). Similarly, N3 discussed how alerting the care team to specific pieces of information could prompt clinicians to investigate certain decisions and to highlight potentially overlooked data to colleagues, which might \textit{``give a bit of confidence to other members of the team.''} (N3).

\qt{N3}{``you could use the vitals to then say actually hold on, this patient doesn't need to be on a ventilator \ldots And then you'd have to then almost give a rationale as to why you're going to continue to ventilate that patient when we know it's better to get them off a ventilator as soon as possible.''}

Another key aspect of collaborative decision-making was to keep family members informed of the clinical decisions \textit{We might invite the sister or senior nurse down to the ward to actually meet that patient, family and go through things} (N2), but also sometimes involve them in the \textit{``clinical decision between the family and the medical team''} (N3).

\subsubsection{Patient Handover Between Clinicians.}
Within the ICU, the patient care team regularly changes due to the nature of working patterns and the need for timely discharge. Participants emphasised the challenges and reduced efficiency resulting from the staffing changes and that \textit{``the changing of clinicians really slows things down''} (N4). The need for familiarisation, in particular, was highlighted as a key factor in delaying clinical decisions as the care team \textit{``want to get to know the patient, make sure everything is well before making that decision.''} (D3). 

\qt{D1}{``one of the biggest challenges is when you come back onto the ward after not being there for some time and you don't know the patient, it can be quite difficult to ascertain exactly what the problem here right now is for a patient who's had a long stay.''}

It follows that effective information transfer during handovers is important, and participants ensured that \textit{``medical and nursing handovers \ldots are structured in that way with the aim of making it as clear as you can.''} (N4). In addition, some frameworks and systems were used to facilitate information transfer during handovers, including \textit{``CareFlow, which would be the handover system where you can access the information about the patient and their investigations, results, their community history''} (N1) and the \textit{``SBAR tool \ldots the Situation, Background, Assessment and Recommendation''} (N4) of the patient. However, handovers often use many different mediums of communication, \textit{``including ICU paperwork, the discharge paperwork that goes with the patient, and CareFlow and Phillips system for receiving information about the patient essentially''} (N1). But this again depends on the context as the outreach team \textit{``don't get a verbal handover of these patients \ldots [but] the ward would have received a verbal handover.''} (N1)

\subsection{Requirements and Capabilities of AI Decision Support Systems}
    
Decisions in the ICU need to be supported by the data captured in computer systems, so explanations and decision support should aid in the interpretation. The framing and level of detail of recommendations were key factors for this. In particular, the high cognitive demand of the ICU requires efficient and simple navigation of these resources, however, facilitating scrutiny of the data is crucial to ensure the validity of decisions suggested, creating a tension between low-cognitive load and detail. This section also reflects on the impact of AI recommendations on decision-making.

\subsubsection{Framing and Presentation of Decision Recommendations.} 

The introduction of AI decision support systems within intensive care raises concerns among clinicians: \qti{D3}{I'm really uncomfortable being recommended an action by the system.} However, clinicians demonstrated an interest in obtaining evidence-based suggestions to help them reach their own conclusions: \qti{D3}{I'd rather be presented the data that's there or things that happen and then let me make that decision.}. N4 highlighted how the acceptance of AI decision support systems will be influenced by how recommendations are framed. For example, participants felt that presenting a recommendation as ``\textit{options available to you}'' would support increased uptake compared to presenting the same recommendation as ``\textit{you should do this}''. Clinicians were also interested in leveraging organ systems in the explanations of an AI decision support system: \textit{``so, it could almost be a flag saying this patient is being held up by brain and cardiovascular''} (D2), with D1 agreeing with the suggestion.

D1 also indicated that lists of supporting factors would be more beneficial than just recommendations as \qti{D1}{if the system is consistently saying `this patient is dischargeable, this patient can go to the ward', I'm not going to trust it in the same way than if it's it uses terminology like `these factors support that this patient is suitable for discharge'.} D3 suggested presenting the recommendations in the form of alerts:

\qt{D3}{``if a computer just would give me an alert\ldots I might be a little bit more prompted to say, oh, there's something wrong, maybe I didn't pick it up, but I wouldn't action it just based on that recommendation.''}

This reliance on evidence and explanations for trust in decisions seems to be integral to the ICU as even if a suggestion (from a human or machine) is wrong, a clinician would use their judgement to reach a conclusion: \qti{D2}{I'm going to note that and respectfully ignore it. Because clinically we feel as a team that... it doesn't represent their clinical status... We do that all the time, so, I don't think it would be a big problem with trust.}

\subsubsection{Complexity and Detail of Explanations.}
A desired feature of an AI decision support system is to provide an appropriate amount of information to clinicians. This would involve avoiding excessive quantity of available data and minimizing cognitive load as \textit{``there’s so much data and information, anything that requires multiple clicks and going between pages if it’s an electronic form, would kind of distract''}(D2). This was seen as clinicians wanting to \textit{``Have a very basic summary''} (D1) and \textit{``some sort of dashboard or some sort of information that’s concisely saying what are the problems here and now that the patient is recovering from''}(D1) as well as \textit{``just clear summaries of what’s gone before, anything that can speed up that process and help clinicians familiarise faster would be helpful to support decision-making''} (N4). Besides having summaries to minimize cognitive load, the number of the sources of information was an issue as \textit{``people get fatigued with all of the different softwares and systems that we use''}(N1), so new systems should, \textit{``be integrated into things we already look at as part of the ward round review.''}
Whilst clinicians want to minimize excessive information, they indicated that \textit{``a layered system would be useful to not overload the user all at once''}(D1), with the option of obtaining more detailed information on demand: \textit{``it’d be great to have the ability to kind of get it, you know click down and get deeper''}(D2).

%\begin{itemize}
%\item The underlying theme is to avoid excessive quantity of available data and minimize cognitive load 
%\item This is captured through the use providing summaries of the patient data as well as integrating any new systems into existing systems to minimize the number of systems present.
%\item Additionally to obtaining summaries, clinicians wanted to obtain more detailed information when required through the use of a layered system with increasing amount of detail.
%\end{itemize}

\subsubsection{Impact of xAI Recommendations on Decision-Making.}
Using a decision support system should inform the actions taken by the clinician. In particular, clinicians know that they may miss important information \textit{``like previous visits and infection statuses and things like that, that in the heat of the moment when you admit a patient, those things can be missed''} (N3) and clinicians acknowledge that they \textit{``are not perfect \ldots [and] forget a lot of option[s]''} (D3). Due to sometimes missing information, clinicians find it useful \textit{``to get alerts from that system [Medway] to let us know those things so we can build up a bit of a picture previously.''}(N3) and \textit{``I wouldn’t see it as an annoyance or a disruption. I would see it as an aid to triggering me to think about something''} (N3). As well as being prompted by the system, clinicians could see the data in different formats as \textit{``a lot of data \ldots [is] easier to see that as a trend''} which can lead to different outcomes produced.
However, whilst clinicians see the benefit of decision-making systems, there are concerns that decision-making systems can lead to obscuring the judgment of the clinicians and the introduction of bias. An example of this is when a decision-making system constrains the number of possible options an individual can take: \textit{``if someone tells me\ldots what about these three options? What if there’s a fourth option \ldots [that's] not presented, and now I’m just thinking about those three options.''}(D3).

%\begin{itemize}
%\item The underlying subtheme of decision xAI recommendations on decision making - outcome based.
%\item The aim would be for the AI system to prompt new thought by presenting miss data or alternative options available to a clinician.
%\item However a concern by the clinicians was that the AI system could obscure the judgement of the clinician and introduce some bias. 
%\end{itemize}

%%%%%%%%%%%%%%%%%%%%%%%%%%
%%                      %%
%%      Discussion      %%
%%                      %%
%%%%%%%%%%%%%%%%%%%%%%%%%%
\section{Discussion}

\subsection{Contextual Nature of ICU Decision-Making}
From our findings, we see that decision-making is difficult due to the highly contextual nature where a decision depends on factors inside and outside of the ICU. This includes operational factors such as resources and staff availability. Whilst these factors are not directly related to the patient's condition, these are important factors that an AI decision-making system would need to take into account. Whilst physiological patient data is important for making decisions, what is considered a `normal' value in the ICU can differ significantly based on the context. This may be because the ICU can safely tolerate more extreme values due to the close monitoring. In addition, the contextual interpretation (e.g., medication, diagnosis, etc.) was highlighted as an important factor in interpreting a biomarker. Therefore, decisions made when examining a biomarker solely without knowing the context may be incorrect and can not be completely trusted. Additionally, rather than using isolated data points for a sense of normality, the past history of a patient's biomarker can give a trend which is important as the trend can indicate if the patient is improving or becoming worse. Therefore, an AI system would need to take into account contextual factors such as past history to aid in supporting decisions. Also, whilst trends in data values can be important, direct examination of visible and behavioural aspects of patients was highlighted as an invaluable component in the decision-making process. Therefore until AI can formalize and represent this as features, it must be acknowledged that such systems do not fully capture the patient state.

\subsection{Human-Introduced Variability}

Patient care within the ICU is largely informed by the observations, experience, and decision-making of individual clinicians. As patients are often managed by multiple care teams, this introduces inherent variability in the information that clinicians have at any point in time, and in how different clinicians perceive the state and requirements of patients. 

Participants highlighted that this variability is particularly relevant in patient handover and collaborative decision-making contexts, in which information on the state of the patient needs to be communicated between clinicians. Participants also highlighted that the need for first-hand observation and validation of a patient's current state by individual clinicians can introduce delays in decisions being made or actioned. In the case of patient handover, participants described the use of the SBAR tool, which aims to mitigate this problem by providing a framework by which clinicians can quickly communicate key information about patients. Similarly, xAI tools that can provide simple overviews and evidenced recommendations relating to individual patients during inter-clinician interactions has the potential to further expedite collaborative processes such as patient handover.

As well as the distributed nature of patient care across multiple care teams, variability between clinicians in the information they have about patients can be caused by missed data. Participants valued AI-generated decision recommendations as an opportunity to consider specific pieces of patient information and alternative decision options that they may not have otherwise considered. However, there was also concern that such recommendations could introduce unwanted bias by unconsciously limiting further thought into alternatives not presented by the system.

\subsection{Design Recommendations}

Based on our findings, we present several recommendations for the design of xAI tools for AI decision support systems in intensive care, and discuss how these can support the integration and acceptance of such tools in clinical practice.

\subsubsection{Minimise cognitive load: integrate into existing systems and include multiple levels of detail with layered explanations.}

As highlighted by several participants, minimizing cognitive load is a crucial factor in the delivery of patient care within ICU. Participants expressed a desire for integration into existing systems and for systems that can provide clear summaries of patient information, with the option of viewing more detailed information when required. For AI decision support tools, explanations should provide succinct summaries of important patient information that can be quickly interpreted by clinicians, as well as the option to view more comprehensive and detailed explanations when needed. Previous work with occupational therapists by \citet{Szymanski2024Daee} highlighted simplified feature importance as an effective way of communicating important patient information, which, when coupled with an additional view of the underlying data, may provide a suitably layered system for use in intensive care that minimises cognitive load.

\subsubsection{Present explanations with additional contextual information.}

It is crucial that explanations of AI clinical decision recommendations are presented in such a way that reflects the highly contextual nature of intensive care. Just as clinicians do not solely rely on physiological data for decision-making, AI decision support systems should not present explanations purely in terms of physiological data. Instead, these explanations should be presented alongside `end-of-bed' information collected through time spent with patients, information concerning the patient's care history and current trajectory, and relevant operational factors such as bed availability.

\subsubsection{Frame explanations in a way that prompts independent thought.}

Participants highlighted that the acceptance of AI decision support systems relies on clinicians feeling confident that these systems are not negatively impacting or biasing their overall decision-making. In particular, participants expressed a desire for \textit{suggestions} rather than required actions, and that any associated explanations should be evidence-based to promote trust in these systems. Designers of xAI tools should consider \textit{how} decisions and their explanations are framed, including the language used, to support clinicians in reaching their own conclusions based on the available information. For example, several participants highlighted how seeing data they may have missed could prompt them to consider options not previously considered.

\subsubsection{Provide tools for communicating patient state to support shared decision-making.}

Our results show that the communication of patient state between clinicians limits efficiency of intensive care decision-making, particularly when handing over patients from one care team to another. Designers should aim to facilitate the clear and efficient communication of patient state, to allow clinicians to quickly build up a picture of patients' care requirements, history, and future care plans, in order to improve handover efficiency. As suggested by participants, one way of achieving this could involve presenting patient information and decision recommendations in formats that better align with existing methods of communication. One example given was to utilise organ systems to present recommendations, matching the thought process and means of communicating clinicians naturally go through when making decisions.

%%%%%%%%%%%%%%%%%%%%%%%%%%%
%%                       %%
%%      Limitations      %%
%%                       %%
%%%%%%%%%%%%%%%%%%%%%%%%%%%
\section{Limitations and Future Work}

While consultants (attending physician-level doctors) provided input during the write-up phase, none participated directly in the interviews due to staffing considerations, limiting the depth of insight from senior medical perspectives. Furthermore, the study was constrained by a single round of interviews, each lasting approximately 35 minutes, offering limited scope for in-depth discussion. To facilitate attendance, interviews were conducted virtually which may have affected engagement, such as with the whiteboard, and interaction between participants during the interviews. Future work could benefit from engaging a broader range of clinicians, including consultants, ensuring in-person attendance, and extending the interview duration or conducting multiple rounds to explore the themes in greater detail.

%%%%%%%%%%%%%%%%%%%%%%%%%%
%%                      %%
%%      Conclusion      %%
%%                      %%
%%%%%%%%%%%%%%%%%%%%%%%%%%
\section{Conclusion}
We present insights gained from interviewing seven intensive care clinicians on the topic of decision-making and systems to support such decisions. Analysis of the resulting transcripts revealed three themes, encompassing the range of factors on which clinicians currently make decisions, considerations for shared decision-making, and requirements of decision support systems. Considering the points raised by the participants of this study, and presenting actionable recommendations will aid developers in the design of AI decision support systems. Better integrating such systems into clinicians' existing decision-making processes will provide improved decision support, facilitating improved patient care whilst minimising additional cognitive load.

\section*{Acknowledgements}
We thank the participants for their input during the workshops. This work was supported by an EPSRC Impact Acceleration Account (EP/X525674/1), EPSRC LEAP Digital Health Hub (EP/X031349/1), and NIHR Invention for Innovation award (AI\_AWARD01943). RSR is funded by a UKRI Turing AI Fellowship (EP/V024817/1).

\bibliographystyle{plainnat}
\bibliography{references}

\appendix

\section{Patient Personas}
\label{app:personas}

\begin{enumerate}
    \item 32 year old female patient admitted three days ago with pyelonephritis. Admitted due to sepsis with persistent hypotension despite adequate IV rehydration on wards and deranged renal function. Clinical improvement – haemodynamically stable and improving biochemistry - following 48 hours of vasopressor support, IV antibiotics and further IV fluids. Currently not requiring any organ support but arterial line and CVC remain in. On assessment, no oxygen requirement, HR 90, BP 115/63, warm peripherally, GCS 15, FB +ve 2.5L LOS with good UO.\\

    \item 68 year old male with background of mild COPD admitted 12 days ago for increasing oxygen requirement due to a community acquired pneumonia. Required intubation on day 2 of admission and remained ventilated for 8 days before being extubated. Yesterday was weaned off oxygen, haemodynamically stable and biochemicaly improving. Today he is pyrexial 37.9 and requiring 8L oxygen. He Looks and feels clinically worse. His respiratory rate is 24, sats 93\% on 8L, HR 98, BP 118/67, GCS 15, passing urine adequately. He had completed his antibiotic course this morning but now has started IV tazocin on microbiology recommendation.
\end{enumerate}

\section{Semi-Structured Interview Questions}
\label{app:questions}

\textit{Background Knowledge Gathering:}
\begin{enumerate}
    \item What information is most important for making a decision for these personas and would this change for other personas?
    \item Are there operational factors or resource constraints which would impact how you make these decisions?
    \item How would time constraints influence how you make these decisions?  
    \item If you were making a similar decision with a colleague, how would you want them to explain their thoughts to you?
    \item How would you explain your thinking to them?
\end{enumerate}

\noindent\textit{Requirements of xAI Systems:}
\begin{enumerate}
    \item How should a system like this explain or back up its suggestions to enable you to best understand its decision?
    \item How much detail would you want explanations to have in different situations? 
    \item How would these explanations support your decision-making process? 
    \item What kind of data would you want to be shown? 
    \item How would you want to view or visualise the data? 
    \item How would explanations effect your trust of the system? 
    \item Would your decision-making be affected if the system disagrees with your decision? If so, how? 
\end{enumerate}

\end{document}